\documentclass[epj]{svjour}
\usepackage{graphics}
\usepackage{epstopdf}
\usepackage[T1]{fontenc}
\usepackage{graphicx}
\usepackage{amsmath,amssymb,slashed,bbm,xcolor}
\DeclareMathAlphabet{\boldmathe}{T1}{cmr}{bx}{it}

\newcommand{\eq}[1]{(\ref{#1})}

\newcommand{\n}[1]{\label{#1}}

\def\beq{\begin{eqnarray}}
\def\eeq{\end{eqnarray}}
\def\ln{\,\mbox{ln}\,}

\def\be{\beta}

\def\de{\delta}

\def\ze{\zeta}

\def\la{\lambda}
\def\na{\nabla}
\def\pa{\partial}

\def\si{\sigma}
\def\om{\omega}
\def\ph{\varphi}

\def\Ga{\Gamma}
\def\De{\Delta}
\def\La{\Lambda}

\begin{document}

\title{From stable to unstable anomaly-induced inflation}


\author{Tib\'{e}rio de Paula Netto\inst{1}\thanks{\emph{E-mail address:} tiberiop@fisica.ufjf.br}, Ana M. Pelinson\inst{2}\thanks{\emph{E-mail address:} ana.pelinson@gmail.com}, Ilya L. Shapiro\inst{1,3}\thanks{\emph{E-mail address:} shapiro@fisica.ufjf.br}, Alexei A. Starobinsky\inst{4,5}\thanks{\emph{E-mail address:} alstar@landau.ac.ru}
%
}                     
%
%
\institute{Departamento de F\'{\i}sica, ICE, Universidade Federal de Juiz de Fora \\
Campus Universit\'{a}rio - Juiz de Fora, 36036-330, MG, Brazil
\and \
Departamento de F\'{\i}sica, CFM, Universidade Federal de Santa Catarina
\\
Bairro da Trindade - Caixa Postal 476, 88040-970, Florian\'{o}polis, SC, Brasil \and \ Tomsk State Pedagogical University and Tomsk State University, Tomsk, Russia
\and \
L. D. Landau Institute for Theoretical Physics RAS, Moscow, 119334,
Russia
\and \
Institute for Theoretical Physics, Department of Physics and Astronomy,
\\ Utrecht University, 3508 TD Utrecht, The Netherlands}

\date{Received: date / Revised version: date}
%









\abstract{ Quantum
effects derived through conformal anomaly lead to an inflationary
model that can be either stable or unstable. The unstable version requires
a large dimensionless coefficient of about $5\times 10^8$ in front of the
$R^2$ term that results in the inflationary regime in the $R+R^2$
(``Starobinsky'') model being a generic intermediate attractor. In this
case the non-local terms in the effective action are practically irrelevant,
and there is a 'graceful exit' to a low curvature matter-like dominated
stage driven by high-frequency oscillations of $R$ -- scalarons, which
later decay to pairs of all particles and antiparticles, with the amount
of primordial scalar (density) perturbations required by observations.
The stable version is a genuine generic attractor, so there is no exit
from it. We discuss a possible transition from stable to unstable phases
of inflation. It is shown that this transition is automatic if the sharp
cut-off approximation is assumed for quantum corrections in the period of
transition. Furthermore, we describe two different quantum mechanisms that
may provide a required large $\,R^2$-term in the transition period.
} 
\authorrunning{T. de P. Netto, A. M. Pelinson, I. Shapiro, A. A. Starobinsky}
\titlerunning{From stable to unstable anomaly-induced inflation}
\maketitle

\section{Introduction}
\label{Sect1}

There are many solid results in Quantum Field Theory (QFT) in
curved space-time, concerning divergences and renormalization
and to the evaluation of finite quantum corrections (see, for
example, \cite{GMM}, \cite{birdav},
\cite{book} and \cite{ParTom} for introduction and further
references, and \cite{Poimpo} for a recent review). The most
interesting applications concern vacuum sector of the theory and
the one-loop approximation is usually considered reliable. Hence
the main interest is usually paid to the quantum effects of free
matter fields on an arbitrary classical gravitational background.
In particular, for the case of free massless conformal fields in a
Friedmann-Lema${\rm \hat i}$tre-Robertson-Walker (FLRW)
isotropic cosmological model, an explicit calculation of the finite
average value of the energy-momentum tensor (EMT) of these fields
is possible by using conformal anomaly \cite{duff77} (see also earlier
pioneer papers \cite{zest71,pf74} on the EMT regularization and
calculation in a more general anisotropic cosmology, and \cite{duff94}
for a general and historical review). The early
works concerning cosmological applications of conformal anomaly
\cite{davies,fhh} led to the first inflationary model \cite{star,star-a},
which was extensively studied (see, e.g., \cite{vile,ander,suen,andsue}
and \cite{hhr}), including inhomogeneous perturbations of this modified
gravity model in the scalar \cite{MuCh,star81,star83} and
tensor~\cite{star81,star83}
sectors (following the pioneer calculation of generation of tensor perturbations
during inflation in the case of the Einstein gravity in \cite{star79}).

The anomaly-induced effective action in $d=4$ was first calculated
in \cite{rie,frts84} (see also \cite{fhh} for the earlier non-covariant
version and \cite{AnMo,a,MaMo} for a more complete local covariant
presentations), similar to the famous Polyakov action
\cite{Polyakov} in $d=2$. The application of this effective action
to cosmology was first considered in \cite{buodsh}, where the possible
torsion terms were also taken into account. Later on, the effective
action approach was systematically pursued in \cite{anju} and
\cite{wave}. The main advantage of using the anomaly-induced
effective action is a better control of approximations and also
better starting point for possible generalizations.

Anomaly-driven inflation can be stable or unstable, depending on
the sign of the local $R^2$-term \cite{star,star-a,MSS-1988}.
If such a term is not introduced at the classical level, the stability
depends on the number of particles of different spin (0, $1/2$
and 1) in the underlying QFT on curved space-time background.
In particular, for the supersymmetric particle content inflation
is stable and, for the
Minimal Standard Model (MSM) of particle Physics, it is unstable
\cite{insusy}. It is possible to have inflation which starts as
stable due to the supersymmetry. After some time supersymmetry
breaks down and inflation becomes unstable. The reason why
supersymmetry can disappear is related to the greater masses
of the $s$-particles that decouple according to the Appelquist
and Carazzone decoupling theorem \cite{AC}. Let us note that the
gravitational version of decoupling theorem has been derived in
\cite{apco}, hence the described scheme looks consistent with
the known QFT results.

A relevant question is why the energy scale of stable inflation is
decreasing, such that the gravitational decoupling could take place.
The solution to this problem has been suggested in \cite{Shocom,asta}.
The stable anomaly-induced inflation is due
to the quantum effects of massless conformal fields, and is
strictly exponential, such that the Hubble parameter is constant.
However, taking the weak effects of masses of quantum fields
into account, one can observe a tempered form of inflation,
with decreasing magnitude of the Hubble parameter.

The second interesting question is what happens with the Universe
after it leaves
the stable inflationary stage. For the choice of parameters which
corresponds to the unstable inflation, there are different types of
solutions \cite{star,star-a}. The desirable one is when the Universe is
asymptotically approaching the FRW-behavior. Then the non-local
part of anomaly-induced action rapidly becomes irrelevant and the
evolution is essentially driven by the local $R^2$ term. Moreover,
in order to control cosmic perturbations after inflation, the
coefficient of this term must be very large, of the order of
$5 \times 10^8$ \cite{star83}. This type of inflationary
model is supported by all known observations, including
Planck data \cite{Ade}.

At the same time, there are other solutions in the theory with
anomaly-induced corrections, which can be called hyperinflation
\cite{anju}. In this case the expansion of the Universe is even
more violent than in the exponential phase, and there is no chance
for a sound physical interpretation of such a solution. The first
purpose of the present work is to see which of the two possible
scenarios of the post-stable evolution takes place. The simplest
possible approach in used. Namely, we assume that the unstable
phase starts exactly at the point where the stable phase ends.
Another important issue discussed in this paper concerns quantum
mechanisms to generate a large coefficient of the $R^2$-term in
the transition period from stable to unstable inflation. We
demonstrate that this effect may take place because of a possible
strong coupling between fields which may result in the large
value of the parameters $\xi$ of the non-minimal interaction
of scalar fields with scalar curvature.

The paper is organized as follows. In order to have a self-consistent
presentation, Sect. \ref{Sect2} includes a brief review of the
effective action induced by anomaly, and also the inflationary
solutions, both stable and unstable, including tempered stable
inflation due to the effects of massive fields. The difficulties
of the QFT-based systematic study of the transition period are
also briefly explained. Sect. \ref{Sect3} describes the numerical
results concerning the transition between stable and unstable
epochs in the sharp cut-off approximation. In Sect. \ref{Sect4}
these results are supported by analysis of the phase diagrams in
both cases. Sect. \ref{Sect5} describes two alternative (but
related) quantum mechanisms to generate a large coefficient of
the $R^2$-term in the transition epoch. Finally, in the last
section we draw our conclusions and discuss further
perspectives of the QFT-based approach to inflation.

\section{Brief review of anomaly-driven inflation}
\label{Sect2}

The effective action of vacuum is defined through the path integral
over the set if all matter fields $\Phi$, including gauge fields and
ghosts (e.g., in Standard Model or GUT's)\footnote{Our notations are
$\eta_{\mu\nu}=diag(+---)$ and
$\,\,R_{\mu\nu}=\partial_\lambda\,\Ga^\la_{\mu\nu}
-\, \cdots \,\,.\,\,$},
\beq
e^{i\Ga(g_{\mu\nu})}\,=\,\int d\Phi \,e^{iS(\Phi,\,g_{\mu\nu})}\,.
\label{path}
\eeq
The consistency of the theory requires that the classical action
includes vacuum part,
\ $S(\Phi,\,g_{\mu\nu})=S_{vacuum}(g_{\mu\nu})
+S_{matter}(\Phi,\,g_{\mu\nu})$, where the first term does not
depend on the matter fields, but still has to be renormalized.
The vacuum action of renormalizable theory should include
Einstein-Hilbert term with a cosmological constant,
\beq
S_{EH}\, =\,
-\,\frac{1}{16\pi G}\,\int d^4x\sqrt{-g}\,(R + 2\La)
\label{EH}
\eeq
and four covariant four-derivative terms,
\beq
S_{HD}\, =\, \int d^4x\sqrt{-g}\,
\left\{ a_1 C^2 + a_2 E + a_3 \Box R + a_4 R^2 \right\}.
\label{HD}
\eeq
Here $\,a_1,..,a_4$ are parameters of the vacuum action.
In the conformal case one can set $a_4=0$, but it is also
possible to include this term. The full action of external
metric is
\beq
S_{vacuum}\, =\, S_{HD}\, + \,S_{EH}\,.
\label{vacuum}
\eeq
Since gravity is not quantized,
there is no problem with unitarity of gravitational $S$-matrix.
Instead, one should worry about the stability of the approximate
low-energy classical solutions below Planck energy scale,
as it was discussed recently in \cite{GWs}.

\subsection{Anomaly and induced effective action}
\label{Sect2.1}

In the very early universe the masses of quantum fields and their
interactions are assumed to be irrelevant. Consider conformal
theory with $N_s$ real scalars, $N_f$ Dirac fermions and
$N_v$ massless vectors. For the scalar massless fields $\ph$
conformal invariance requires that the nonminimal parameters
of the $\xi R\ph^2$-interaction are $\xi=1/6$. Taking $a_4=0$,
the action $S_{HD}$ satisfies the conformal Noether identity
\beq
- \, \frac{2}{\sqrt{-g}}\,g_{\mu\nu}
\frac{\de S_{HD}}{\de g_{\mu\nu}}\, = \, 0\,,
\label{identity1}
\eeq
which means zero trace for the stress tensor of vacuum $T^\mu_\mu=0$.
At the quantum level, this condition is violated by anomaly,
\beq \label{main eq}
&&T\,\,=\,\,
\langle T_\mu^\mu \rangle\,=\,- \frac{2}{\sqrt{-g}}\,g_{\mu\nu}
\frac{\de {\bar \Ga}^{(1)}}{\de g_{\mu\nu}}
\\
&=&
- \,(wC^2 + bE + c \Box R)\,.
\nonumber
\eeq
where $\,w$, $\,b\,$ and $\,c\,$ are $\beta$-functions for
the parameters $\,a_1,a_2,a_3$, which depend on the number
of (real) scalar, (Dirac) spinor and gauge vector fields,
$\,N_s$, $N_f$, $N_v$,
\beq
w \,=\, \be_1\,=\,\frac{1}{(4\pi)^2}\,\Big(
\frac{N_s}{120} + \frac{N_f}{20} + \frac{N_v}{10}\Big)\,,
\label{w}
\eeq
\beq
b \,=\, \be_2\,=\, -\,\frac{1}{(4\pi)^2}\,\Big( \frac{N_s}{360}
+ \frac{11\,N_f}{360} + \frac{31\,N_v}{180}\Big)\,,
\label{b}
\eeq
\beq
c \,=\, \be_3\,=\,
\frac{1}{(4\pi)^2}\,\Big( \frac{N_s}{180} + \frac{N_f}{30}
- \frac{N_v}{10}\Big) \,.
\label{c}
\eeq

One has to note that the coefficient $c$ has the well-known
regularization-dependent ambiguity, which is equivalent to
the possibility to add the $a_4 R^2$-term at the classical
level (see, e.g., \cite{duff94}). This issue was discussed
in full details in \cite{anom-03}, using both dimensional
and covariant Pauli-Villars regularization. It was shown
that the ambiguity concerns the starting point of the
renormalization group trajectory and not the flow itself.
In particular, this means one can fix it by imposing a
renormalization condition on the classical coefficient $a_4$.
There is nothing wrong in defining the $R^2$ term by hand,
but it is more natural to assume that the this term comes from
vacuum quantum effects according to (\ref{main eq}), that
corresponds to the point-splitting \cite{chr} and $\ze$
regularizations \cite{dow} \footnote{The equivalent n-wave and
adiabatic EMT regularizations proposed earlier in \cite{zest71}
and \cite{pf74} respectively lead to the same result for conformal
anomaly if applied to the case of a non-zero rest mass $m$ of a
quantum field with $m$ set to zero in the final result.}.
In Sect. \ref{Sect5} we shall
discuss the possibility of significant change of the overall
coefficient  $a_4$ of the $R^2$-term in the epoch of transition
from stable to unstable inflation.

Natural question concerns possible effect of higher loops.
Let us remember that the nonperturbative structure of conformal
anomaly is basically the same as at one loop. This statement is
known as $a$- and $c$-theorems and gained significant attention
in the recent years \cite{Koma,Polchi}. \  Since only the trace anomaly
is relevant for the dynamics of conformal factor, one can safely
assume that at higher loops nothing changes dramatically and
conclude that the one-loop approximation is sufficiently
reliable in this case.

The one-loop part ${\bar \Ga}^{(1)}$ of the vacuum effective
action satisfies the equation
\beq
- \frac{2}{\sqrt{-g}}\,g_{\mu\nu}
\frac{\de {\bar \Ga}^{(1)}}{\de g_{\mu\nu}}
\,=\,\langle T_\mu^\mu \rangle\,,
\label{main equation}
\eeq
which can be solved in the form \cite{rie,frts84}
\beq
\nonumber
\Ga_{ind}
&=&
S_c[{\bar g}_{\mu\nu}]
\\
\nonumber
&+&
\int d^4 x\sqrt{-{\bar g}}\,\Big\{
\,w\si {\bar C}^2 + b\si \big({\bar E}-\frac23 \bar{\bar \Box}
{\bar R}\big)
+ 2b\si{\bar \De}_4\si
\\
&-&
\frac{1}{12}\,\big(c+\frac23 b\big)
\big[{\bar R}
- 6({\bar \na}\si)^2 - 6 \bar{\bar \Box}\si \big]^2\Big\}\,,
\label{quantum1}
\eeq
where we separated the conformal degree of freedom $\si$
according to
\beq
g_{\mu\nu}
 &=&
{\bar g}_{\mu\nu}\cdot e^{2\si(x)}
\,=\,
{\bar g}_{\mu\nu}\cdot a^2(x)
\label{asigma}
\eeq
and used notation
\beq
\De_4 = \Box^2 + 2\,R^{\mu\nu}\na_\mu\na_\nu - \frac23\,R\Box
+ \frac13\,(\na^\mu R)\na_\mu
\eeq
for the fourth derivative, conformal operator acting on
conformal-invariant scalars. The term $S_c$ in (\ref{quantum1})
is a conformal invariant functional,
 $\,S_c[{\bar g}_{\mu\nu}]=S_c[g_{\mu\nu}]\,$
which is an ``integration constant'' for the
equation (\ref{main equation}). In cosmology, this term is
irrelevant for defining the dynamics of the conformal factor
of the background metric, $a(\eta)$ and therefore (\ref{quantum1})
is the exact form of quantum correction in this case.
The general fiducial metric is
\beq
\label{metric}
d{\bar s}^2\,=\,{\bar g}_{\mu\nu}dx^\mu dx^\nu
= d\eta^2 - \frac{dr^2}{1-kr^2}-r^2 d\Omega\,,
\label{flat1}
\eeq
where $\eta$ is the conformal time. In what follows
we consider spatially flat metric, $k=0$.

\subsection{Stable and unstable solutions}
\label{Sect2.2}

The dynamics of conformal factor is defined from the variational
principle of the total action, including quantum corrections,
\beq
S_t=S_{vacuum}+\Ga_{ind} \,.
\label{massless}
\eeq
Then we arrive at the following equation:
\beq
\frac{{\stackrel{....}{a}}}{a}
+\frac{{3\stackrel{.}{a}} {\stackrel{...}{a}}}{a^2}
+\frac{{\stackrel{..}{a}}^{2}}{a^{2}}
-\left( 5+\frac{4b}{c}\right)
\frac{{\stackrel{..}{a}} {\stackrel{.}{a}}^{2}}{a^3}
\frac{{\stackrel{..}{a}}}{a^{3}}
\nonumber
\\
-\frac{M_{P}^{2}}{8\pi c}
\left( \frac{{\stackrel{..}{a}}}{a}+
\frac{{\stackrel{.}{a}}^{2}}{a^{2}}
-\frac{2\Lambda }{3}\right)
\,=\,0\,,
\label{foe}
\eeq
where $\,M_P^2=1/G\,$ is the square of the Planck mass.
We assume that the cosmological constant  $\La$ always satisfies
the condition $0 < \La \ll M_P^2$. Eq. (\ref{foe}) is written in
terms of the physical time
$t$, where $\,dt=a(\eta)d\eta$. An equivalent third-order
equation (\ref{T00}) can be obtained as $00$-component of
Einstein equation with quantum corrections \cite{fhh}.
Moreover, this equation can be reduced to a first-order one,
as shown in Sect. \ref{Sect4}.
Let us note that more detailed discussion of deriving 00- and
ij-components of generalized Friedmann equations was given in
\cite{hhr} and generalizations in the presence of radiation
in \cite{RadiAna}.

The equation (\ref{foe}) has important particular solutions
with constant Hubble parameter \cite{star,star-a} (for $\La\neq 0$
the solution was obtained in \cite{asta}),
\beq
a(t) &=& a_0 \cdot e^{H_0t}\,,
\label{flat solution}
\eeq
where
\beq
H_0=H_{\pm}=\frac{M_P}{\sqrt{-32\pi b}} \cdot
\Big(1\pm
\sqrt{1+\frac{64\pi b}{3}\frac{\La}{M_P^2}}
\,\,\Big)^{1/2}_{.}
\label{H}
\eeq
The $H_{+}$ solution is close to the original one of
\cite{star,mamo}, \ $H={M_P}/\sqrt{-16\pi b}$, which is an
exponential inflationary solution. The second value is close
to the classical dS solution $\,H_- \approx \sqrt{\Lambda/3}\,$
without quantum corrections.  In what follows we will mainly
concern the inflationary phase and, therefore, assume
$\,H_0=H_+$.

The solution (\ref{H}) is real since $b<0$, according to (\ref{b}).
At the same time, the coefficient
$c$ in Eq. (\ref{c}) may have different signs, depending on
the particle content of the theory. For small perturbations
\ $\si(t) \to \si(t) + \de \si(t)$ \ around inflationary
exponential solution, it is easy to check that it is stable
for $c>0$ and unstable for $c<0$ \cite{star,asta}.
Assuming (\ref{c}),  the stability condition  $c>0$ boils
down to the relation
\beq
\frac13\,N_f\,+\,\frac{1}{18}\,N_s \,>\,N_v \,,
\label{condition}
\eeq
satisfied for any realistic supersymmetric model.
Then $H=H_{+}$ is the unique stable attractor and
hence inflation starts for any choice of initial conditions with
homogeneous and isotropic metric.

On the other hand, (\ref{condition}) is not satisfied for
the Minimal Standard Model of Particle Physics \cite{Shocom}.
Another case when the condition (\ref{condition}) is not
satisfied is the present-days universe. Since the decoupling of
the lightest  massive particles (presumably neutrino), photon is
the  unique ``active'' quantum particle, such that
\ $N_v = 1$ \ and \ $N_f=N_s=0$.

Let us note that in this case the classical dS solution
with  $\,H=\sqrt{\Lambda/3}\,$ is stable under small perturbations
of Hubble parameter \cite{asta}, which is a relevant consistency
test for the whole approach. The same is true for the tensor
perturbations, which do not grow up for dS \cite{star79,wave,hhr}
and for other classical solutions, even in the presence of matter
or radiation \cite{GWs}.

\subsection{Effect of masses and tempered stable inflation}
\label{Sect2.3}

Consider some realistic supersymmetric model, where $s$-particles
have relatively large masses. Other particles can be approximately
regarded as massless.

At the beginning of the stable inflation Hubble parameter $H$
is even greater that all masses and the last can be seen as small
perturbations violating conformal invariance. In this case one
can apply a conformal description of the massive theory
\cite{cosmon} (similar approach can be found in \cite{Percacci}).
The masses of matter fields, Newton constant and cosmological
constant are replaced by powers of a new auxiliary scalar $\,\chi$,
\beq
\nonumber
m_s^2 &\to& \frac{m_s^2}{M^2}\,\chi^2\,,
\nonumber
\\
m_f &\to& \frac{m_f}{M}\,\chi\,,
\\
\frac{1}{16\pi G}\,R &\to& \frac{M_P^2}{16\pi M^2}\,
\left[\,R\chi^2 + 6\,(\pa \chi)^2\,\right]\,,
\nonumber
\\
\La &\to&\frac{\La}{M^2}\,\chi^2 \,,
\label{replace}
\eeq
where $\,M\,$ is a new dimensional parameter.
For the Einstein-Hilbert term the kinetic term for $\chi$
provides conformal invariance of the action. In order to have
local conformal invariance one can define that the field
$\chi$ transforms as
\beq
\chi \to \chi\,e^{-\si(x)}\,,
\label{auxiliar}
\eeq
while other fields transform according to
\beq
\nonumber
g_{\mu\nu}\to g_{\mu\nu}\cdot e^{2\si(x)}\,,
\nonumber
\\
\varphi \to \varphi \cdot e^{-\si(x)}\,,
\\
\psi \to \psi\cdot e^{- 3\si(x)/2}.
\eeq
Now we can calculate anomaly and anomaly-induced effective
action.   Finally, we fix the conformal gauge according to
$\,\chi = {\bar \chi}\,e^{-\si}=M$. The result has the form
\beq
\Gamma^{(1)} &=& S_{vacuum}\,+\,\Ga_{ind}
 \nonumber \\ &-&
\int d^4 x\sqrt{-{\bar g}}
\,e^{2\si}\,\big[{\bar R}+6({\bar \na}\si)^2\big]
\,\cdot\,\Big(\, \frac{1}{16\pi G} - f\cdot\si\,\Big)
\nonumber
\\
&-&
 \int d^4 x\sqrt{-{\bar g}}\,e^{4\si}\,\cdot\,
\Big(\frac{\La}{8\pi G}\,-\,g\cdot\si\,\Big)\,,
\label{qfm}
\eeq
where the ``massless'' terms in the {\it r.h.s} were defined
in (\ref{vacuum}) and (\ref{quantum1}) and the coefficients
$f$ and $g$ can be expressed via the dimensional parameters
\beq
\label{ftilde}
\tilde{f} &=&
\frac{16\pi\,f }{M_P^2}
\\
&=&
\frac{1}{2\pi}\,\sum_{scalars}\,\frac{N_s\,m_s^2}{M_P^2}
\,\Big(\xi - \frac16\Big)
\,+\,
\frac{1}{3\pi}\,\sum_{fermions}\,\frac{N_f\,m_f^2}{M_P^2}
\,,
\nonumber
\\
\label{replace11}
\tilde{g} &=& \frac{8\pi\,g}{M_P^2\La}
\\
&=&\frac{1}{4\pi}\,\sum_{scalars}\,\frac{N_s\,m_s^4}{M_P^2\La}
\,-\,\frac{1}{\pi}\,\sum_{fermions}\,\frac{N_f\,m_f^4}{M_P^2\La}\,.
\nonumber
\eeq
In the last expression we assume for simplicity that masses of all
fermions are equal, the same with scalars masses and non-minimal
parameter $\xi$. The possibility of $\xi \neq 1/6$ is introduced
for generality, more detailed discussion will be given in Sect.
\ref{Sect5}.

The expression (\ref{qfm}) is not an exact result like
(\ref{quantum1}), even for the FRW metric. The reason is that the
conformal invariant functional $S_c$ in this case depends not
only on the metric  $g_{\mu\nu}$, but also on the scalar field
$\chi$. The approximation which provides Eq. (\ref{qfm}) becomes
clear if we remember that the renormalization group in
curved space is related to the global rescaling of the metric,
\ $g_{\mu\nu} \to g_{\mu\nu}\cdot \exp(2\tau)$, \
\cite{nelspan82,buch84,Toms83,book}.
Since coefficients $\om,\,b,\,c,\,f,\,g$ are the Minimal Subtraction
scheme - based $\be$-functions of the higher derivative parameters,
$G^{-1}$ and $\rho_\La=\La/8\pi G$, it is easy to note that
(\ref{qfm}) is exactly the renormalization group improved classical
action of vacuum (\ref{vacuum}), where the global scaling parameter
is replaced by the time-dependent conformal factor of the metric,
\ $\tau \to \si(t)$. Hence the approximation assumed in (\ref{qfm})
is the one of the Minimal Subtraction scheme of renormalization.
Within this approximation one can not observe effects of masses,
such as low-energy decoupling. However, it is a reliable
approximation at high energies, including at the initial
stage of the stable anomaly-driven inflation \footnote{This
result of Refs. \cite{Shocom,asta} concerning the effects
of massive fields has been independently confirmed in \cite{PaGo}
by technically different method (see also \cite{Percacci}).}.

Different from the effective action of massless fields \cite{rie,a},
the covariant version of Eq. (\ref{qfm}) is not known,
but this expression is sufficient for basic cosmological application.
One can safely assume that the cosmological constant
and its running do not play essential role at the inflation
epoch. Following \cite{Shocom,asta}, we set $\La=\tilde{g}=0$.
Then the equation for  $\si(t)$ is
\beq
\n{eqmotion}
\stackrel{....}{\si}
+ 7 \stackrel{...}{\si} \dot{\si}
+ 4 \ddot{\si}^2
+  4 \Big(3 - \frac{b}{c} \Big) \ddot{\si} \dot{\si}^2
- \frac{4b}{c}\dot{\si}^4
\nonumber
\\
- \frac{M^2_p}{8\pi c}
\Big[ \left(\ddot{\si} + 2\dot{\si}^2 \right)
\big(1 - \tilde{f}\si\big)
- \frac{\tilde{f}}{2} \dot{\si}^2 \Big]\,=\,0\,.
\eeq
The new part compared to Eq. (\ref{foe}) is the presence of
the mass-dependent terms with $\tilde{f}$, also we use other
variable, according to Eq. (\ref{asigma}).

An approximate solution of Eq. (\ref{eqmotion}) can be
obtained by the replacement
\beq
M_P^2
&\longrightarrow&
M_P^2\,\Big[1 - \tilde{f}\,\si(t)\Big]
\label{replace1}
\eeq
in the expression for the Hubble parameter (\ref{H})
corresponding to the massless solutions (\ref{flat solution}).
The solution has the form
\beq
\si(t) &=& H_0t\,-\,\frac{H^2_0}{4}\,\tilde{f}\,t^2\,,
\label{parabola}
\eeq
that reproduce numerical solutions of (\ref{eqmotion}) with a
very good precision.
The formula (\ref{parabola}) describes a tempered form of
inflation, which starts as en exponential (massless-based)
version (\ref{flat solution}) and ends when $H$ decreases
to the value where the decoupling of $s$-particles starts.

Starting from the
solution (\ref{parabola}), we need to know when the stable
phase of inflation ends and what happens with the universe
after that. If $H^*$ is the energy scale where the supersymmetry
breaks down, the stability holds until the moment of time \ $t_*$,
when \ $H(t_*) = H_0 - (1/2)H_0^2\tilde{f}t_*=H^*$. This
expression means that at the scale $H^*$ most of $s$-particles
are beyond the IR cut-off, $M_s > H^*$  and decouple
from gravity at the quantum level. After certain amount of
such particles decouple, the sign of the $\be$-function
$c=\be_3$ in (\ref{c}) changes to the opposite and inflation
gets unstable. It is a natural to suppose that \ $H^*$ \ should
have the same magnitude as the mass scale of supersymmetry breaking,
$M_{SUSY}$. Another quantity which depends on the same scale is
$\tilde{f} \sim (H^*/M_P)^2$.

The next issue to address
is what happens to the universe after it goes
through the critical point \ $H(t)=H^*$. Indeed, this question is
very difficult to answer and hence we have to change the style of
the consideration. Until this moment our consideration was based
on the use of the QFT results, such as Eq. (\ref{eqmotion}),
even if the derivation of the equation required some risky methods
such as conformal replacement of dimensional parameters \cite{Shocom}.
Unfortunately, in the vicinity of the
critical point $H=H^*$ the QFT provides no real help and even
no insight. The reason is that we do not have approximation or
approach to deal with the situation when the Hubble parameter
is of the same order of magnitude as the mass of the free
quantum field on curved background. It is obvious that the
usual expansion in powers of curvature tensor over square
of mass of the quantum field has no much sense.
On the other hand, the approach which led us to (\ref{eqmotion})
is based on treating mass as a small perturbation and, therefore,
also does not work for $H(t)\sim H^*$.

In the next two sections we will deal with this difficult problem
in a most simple way, which can be called the ``sharp cut-off''
method. Namely, we consider the final point of the evolution
(\ref{parabola}) as a starting point of the unstable inflation.
Of course, this is a great simplification, since we completely
ignore the quantum effects in the intermediate epoch. In the
``sharp cut-off'' approach we simply cut-off the intermediate-scale
quantum effects and try to see in which part of the phase diagram
of the unstable inflation we arrive by moving along the
solution (\ref{parabola}).

\section{Numerical study of transition to unstable epoch}
\label{Sect3}

Consider that the stable phase ends at the moment $t_*$ when
the Hubble parameter has the value approximately corresponding
to (\ref{replace11}),
\beq
H^* = H(t_*) = \sqrt{\tilde{f}} M_P\,.
\label{H*}
\eeq
Using the last equation and Eq. (\ref{parabola}), the value of
$t_*$ can be expressed as
\beq
\n{tstar}
t_* &=&
\frac{2}{H_0^2 \tilde{f}} \Big( H_0 - \sqrt{\tilde{f}} M_P \Big)\,.
\eeq
The initial conditions for the consequent unstable evolution can be
calculated through Eqs. \eq{tstar} and \eq{parabola}, using the Minimal
Supersymmetric Standard Model (MSSM) particle content,
$N_{s,f,v} =(104, 32, 12)$. We assume that $\tilde{f}=10^{-4}$,
which is close to the value for the GUT-scale supersymmetry.

An explanation of the choice of the value of
$\tilde{f}$ is in order here. The GUT-scale supersymmetry is
different from the usual $TeV$-scale supersymmetry which is
useful to address the gauge hierarchy problem. The unique aspect
of supersymmetry which is relevant for us here is the change of
sign in the relation (\ref{condition}) when the supersymmetry is
broken, hence we are not confined by some specific scale of
supersymmetry breaking. Looking at the definition (\ref{ftilde})
it is clear that the value corresponding to the GUT-scale
supersymmetry should be $\tilde{f} \propto 10^{-6}$ and for
the $TeV$-scale supersymmetry about $\tilde{f} \propto 10^{-32}$.
Let us  note that an inflationary model that pretends to describe
the  observed power spectrum of density perturbations (without
inflaton), then it should have $H \approx 10^{14}\,GeV$.
Then ${\tilde f}$ should be no less than $10^{-10}$. Since for
the GUT scale SUSY one has ${\tilde f} \approx 10^-6$ and this
is the mostly interesting case. We have checked numerically that
the qualitative aspects of transition which will be discussed below
are not sensible to the choice of $\tilde{f}$. For making plot and
presentation in general better, we mainly use larger value of
$\tilde{f} = 10^{-4}$. This value also exceeds the value
$\tilde f\approx 10^{-5}$ at which the comoving scale
corresponding to the present Hubble radius first crossed
it in the opposite direction during unstable inflation in
the $R+R^2$ model, see Sect. \ref{Sect5}  below.

Let us present the results of the numerical solution of the equation
(\ref{foe}) in the unstable case with the initial data corresponding
to the point $H^*$. For the sake of  definitiveness, we consider the
Minimum Standard Model (MSM) particle content $N_{s,f,v} =(4, 24, 12)$.

\begin{figure}[!h]
\begin{center}
\includegraphics[height= 6.0 cm,width=\columnwidth]{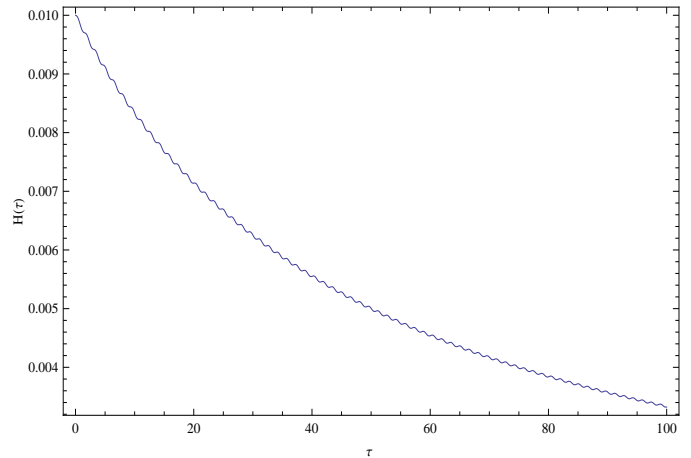}
\caption{Numerical solution for the Hubble parameter $H(t)$ in the
units of Planck time $\tau \equiv  t/t_P$ for the MSM particle
content and $\tilde{f} = 10^{-4}$.}
\label{fig2}
\end{center}
\end{figure}

\begin{figure}[!h]
\begin{center}
\includegraphics[height= 6.0 cm,width=\columnwidth]{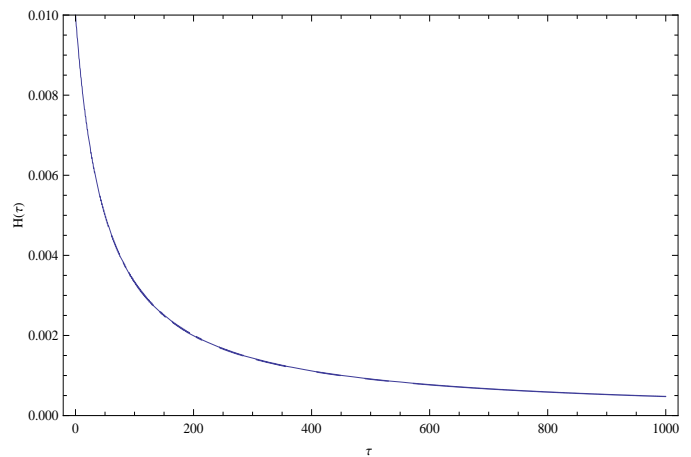}
\caption{The same case as in Fig. \ref{fig2}, but at
larger scale. MSM and
$\tilde{f} = 10^{-4}$.}
\label{fig3}
\end{center}
\end{figure}

The numerical solution at small scales of time show oscillations
in the Hubble parameter $H$ as it is shown in Fig. \ref{fig2}.
\\
\\
Later on amplitude of these oscillations becomes smaller
and the Hubble parameter behavior start to looks very
similar to the radiation-dominated universe without quantum
corrections $H(t) \sim 1/2t$. The situation is illustrated
in Fig. \ref{fig2} and Fig. \ref{fig3}.
\\

\begin{figure}[!h]
\centering
\includegraphics[height= 6.0 cm,width=\columnwidth]{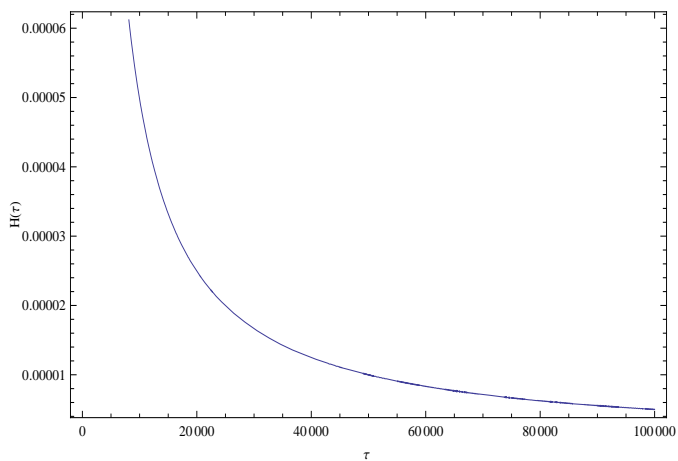}
\caption{Numerical solution for the Hubble parameter $H$ in units of
Planck time $\tau \equiv  t/t_P$ for the photon case
and $\tilde{f} = 10^{-4}$.}
\label{fig5}
\end{figure}

The oscillations which we observe in these plots can lead to the
production of matter particles. But since the physical unstable
inflation is still to come, this is not a physically relevant
process. After certain period corresponding to the MSM, the
expansion of the universe becomes weaker and at some point even
the contributions of massive non-supersymmetric particles get
decoupled. At the last stage only the massless particle - photon,
gives contributions.
The numerical analysis for the unstable inflation driven by a
single photon as an active quantum field, $N_{s,f,v} =(0,0,1)$
leads to the plots shown in Fig. \ref{fig5}.
In this case one can observe, once again, some oscillations
for the initial short period of time, however the amplitude
of the oscillations is weaker compared to the MSM case.
And oscillations become weaker after the initial period.
\\

The case of a single photon is an extra example of stabilizing
perturbations. For the reasons explained above, it is not useful
for describing inflation, and is included mainly for generality.
At the same time, it has some physical relevance, not linked to
inflation, but serving as a test for the consistency  of the
anomaly-induced model. Since we are dealing with the higher
derivative action, it is important to ensure that the physically
relevant solutions do not suffer from the Ostrogradsky-like
instabilities. Looking backward to the history of the universe,
the last strong perturbation for the conformal factor occured
at the epoch when quantum contributions of neutrino decouple
from gravity. According to the plot of Fig. \ref{fig5} the
perturbations for the conformal factor are stabilized
after this decoupling.

\begin{figure}[!h]
\centering
\includegraphics[height= 6.0 cm,width=\columnwidth]{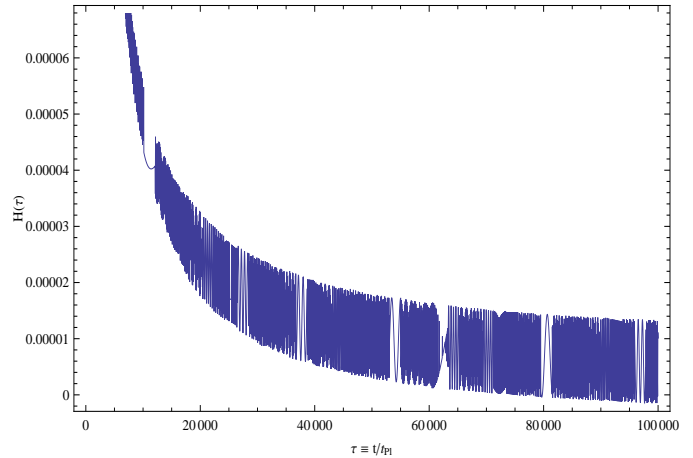}
\caption{
Part of the plot from Fig. \ref{fig2} and Fig. \ref{fig3}, but with
much smaller scale in the $H$ axis.}
\label{fig5a}
\end{figure}

The difference between the photon case and the one of
MSM is quantitative, namely oscillations have smaller
amplitudes for a photon. In order to show that the
oscillations still take place, we show a part of the
previous plots with other scale in Fig. \ref{fig5a}.
\\

\section{Phase diagrams and stitching the solutions}
\label{Sect4}

In this section we shall consider the transition between stable
and unstable regimes by means of phase diagrams. Instead of Eq.
(\ref{parabola}) one can use $00$-component of the equation
\beq
R_{\mu\nu}\,-\,\frac{1}{2}\,g_{\mu\nu} (R-2\La) &=&
8\pi G\,\langle T_{\mu\nu} \rangle\,,
\label{Tmn}
\eeq
in our case it has the form of the third-order equation
\beq
\label{T00}
\frac{2\, {\stackrel{.}{a}}\, {\stackrel{...}{a}} }{a^2}
-\frac{{\stackrel{..}{a}}^2}{a^2}+
\frac{2\, {\stackrel{..}{a}} \, {\stackrel{.}{a}}^2}{a^3}
-\Big(3+\frac{2 b}{c}\Big)\frac{{\stackrel{.}{a}}^4}{a^4}
-\frac{M_P^2}{8\pi c} \, \frac{{\stackrel{.}{a}}^2}{a^2}
\,=\, 0\,.
\eeq
The Eq. \eq{T00} can be reduced to the first order differential
equation
\beq
\label{phase}
\frac{dy}{dx}
&=&
\frac{b(x-x^{-1/3})}{6 c y} - 1
\eeq
by the following change of variables \cite{star,star-a}:
\beq
\nonumber
\label{x}
x &= &\Big(\frac{H}{H_0}\Big)^{3/2}
\,,\\
\nonumber
y &=& \frac{\dot{H}}{H_0^{3/2}} \, H^{-1/2}
\,,\\
dt &=& \frac{dx}{3 H_0 \,x^{2/3} \,y}\,,
\eeq
where (as before) $H_0={M_P}/\sqrt{-16\pi b}$.

\begin{figure}[!t]
\centering
\includegraphics[height= 6.0 cm,width=\columnwidth]{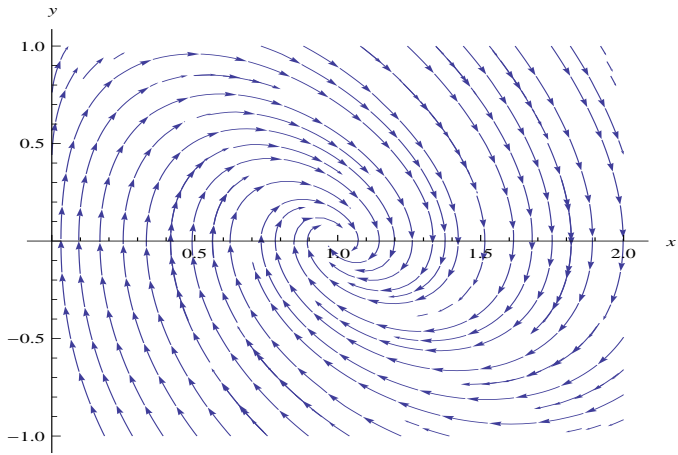}
\caption{Phase diagram of Eq. \eq{phase} with MSSM particle content.}
\label{fig6}
\end{figure}

The critical point $(1,0)$ corresponds to the exponential solution
\eq{flat solution} in both stable and unstable cases. For a stable
inflation \cite{asta} based on the MSSM particle content, the phase
diagram of Eq. \eq{phase} is shown in Fig. \ref{fig6}.
As we can see, there is a single attractor corresponding to the
inflationary solution \eq{flat solution}.
\\

\begin{figure}[!h]
\centering
\includegraphics[height= 6.0 cm,width=\columnwidth]{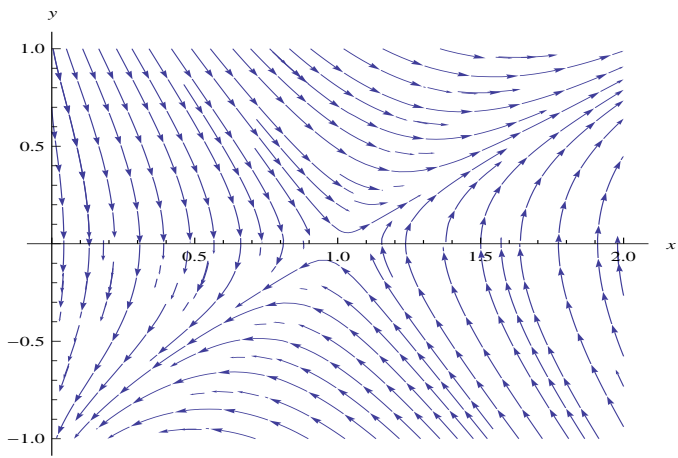}
\caption{Phase diagram of Eq.\eq{phase} with MSM particle content.}
\label{fig7}
\end{figure}

For the unstable case we arrive at the phase diagram shown in
Fig. \ref{fig7}. In this case there are different attractors
\cite{star,star-a}.
\\

Let us see which integral curve in the unstable case of
Fig. \ref{fig7} corresponds to the initial point $x_0$, $y_0$
where the stable regime ends. Replacing the solution
(\ref{parabola}) into expressions \eq{x} we find
\beq
\label{x0}
x_0 &=& (-16 \pi b \tilde{f})^{3/4}
\,,\qquad
y_0 \,=\, -\,\frac{1}{4}\,
\left(\frac{\tilde{f}^3}{-16 \pi b}\right)^{1/4}\,.
\eeq

\begin{figure}[!h]
\centering
\includegraphics[height= 6.0 cm,width=\columnwidth]{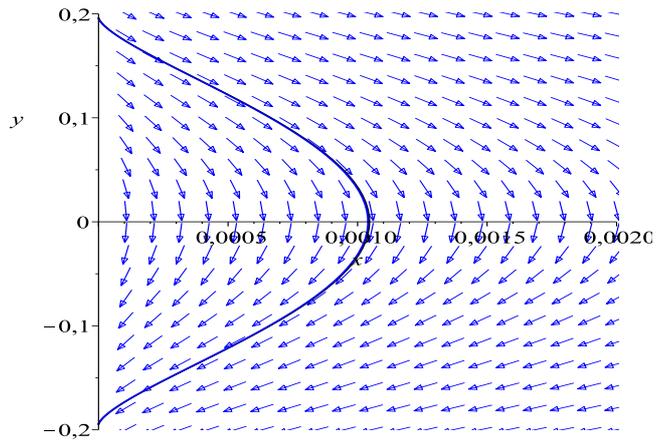}
\caption{Integral curve corresponding to Eqs. \eq{x0}
\ and \  $\tilde{f}=10^{-4}$.}
\label{fig8}
\end{figure}

Taking this initial value, one meets the integral curve shown in
Fig. \ref{fig8}.
\\

Let us say that this result nicely fits our most optimistic expectations.
It exactly corresponds to the relatively ``peaceful'' unstable inflation
qualitatively similar to the one of \cite{star,star-a}, and not to the
``hiperinflation''-type explosion, which was described in  \cite{anju}.
 Anyway, the initial phase of such an inflation can be very violent,
 because the coefficient of the $R^2$ is very small. In the next section
we describe
how it can be enormously increased during the transition
period, which was simply ignored until now within the
``sharp cut-off'' simplification.

\section{UV/IR running and generating huge $R^2$-term}
\label{Sect5}

As we have seen in the previous sections, the end of the stable
phase of inflation occurs in the region of the phase plane from
which the universe can continue into the unstable phase.
However, in order to have a successful inflationary model we
need to go beyond the anomaly-induced effective action.
The reason is that the coefficient of the overall $R^2$-term
with a much greater value, about $5 \times 10^8$, is requested to
control density perturbations after the inflation period ends
\cite{star83}.
In more details: the dimensionless coefficient in the action in
front of the $R^2$ should be $\,N^2/(288\pi^2A_s)$,
where $A_s(k)$ is the amplitude of the power spectrum of primordial
scalar (density) perturbations, while $N$ is the number of $e$-folds
from the end of inflation and $\log(k_{fin}/k)$ at the same
time\footnote{$A_s$ is related to the quantity $A$ used in the
paper \cite{star83} as $A_s=\frac{A^2}{8\pi^2}$.}. $A_s(k)$ is
also proportional to $N^2$ for the model involved.
According to the most recent measurements \cite{Ade}
$\,A_s(k)\approx 2.2 \times 10^{-9}$ for $k=k_0=0.05\,Mpc^{-1}$.
Choosing $N=55$ for $k=k_0$ we arrive at the estimate
$5 \times 10^8$ for the coefficient of the $R^2$-term.
Thus, it is the observed smallness of large-scale inhomogeneous
perturbations in the present Universe (characterized by the small
value of $A_s$) that requires the coefficient in front of the $R^2$
term to be large and of the order of $A_s^{-1}$ during the last,
unstable and observable part of inflation. An alternative to this
could be to add some other non-gravitational scalar field by hand
which would support the second (unstable) part of inflation. The
simplest models of such kind of double inflation were investigated
in \cite{KLS85,GMS91}. However, both models considered in
these papers use trans-Planckian values of this inflaton field and
produce too large amount of primordial gravitational waves which
has been excluded by recent observational data \cite{Ade}.

Note that the same observational data on the power spectrum of
primordial density perturbations in the Universe show also that
any higher order terms of the type $R^n, n>2$ added to the action
of the unstable anomaly-induced inflation are strongly
(exponentially in $n$) suppressed for the number of e-folds from
the end of inflation $N\lesssim 60$ \cite{Huang14}.
Thus, observations demand the absence of significant higher-order in
$R$ corrections to the phenomenological $R+R^2$ inflationary model that
provides an independent support to theoretical arguments for the
conservation of the structure of the conformal anomaly in higher loops
discussed in Sect. \ref{Sect2.1}.

Quantum decoupling of $s$-particles can explain
the chan\-ge of the sign of $c$, but can not make it grow so much.
The purpose of the present section is to discuss alternative
mechanisms which can produce a dramatic change of the
fcoefficient $c$ in the epoch close to the change of its
sign\footnote{Recently another mechanism of generating a
sufficiently large value of $a_4$ was discussed in
\cite{Kawai} in the models with extra dimensions.}.
Let us stress that we have no reliable information about the
physical theories at the GUT scale or even the supersymmetry
breaking scale when the transition from stable to unstable
versions of anomaly-driven inflation is supposed to occur. Hence
we are not in a position to indicate a definite mechanism which
provides such a dramatic growth of the coefficient of
the $R^2$-term. Instead we shall describe two possible situations
when such a growth is possible. In both cases the consideration
is based on the relation between non-minimal scalar-curvature
interaction and vacuum $R^2$-term. This relation was
previously discussed, e.g. in the context of supersymmetry
\cite{KetSta}.

\subsection{RG running of the non-minimal parameter $\xi$}

Renormalization group (RG) in curved space (see, e.g., \cite{book})
tells us that the values of all parameters of the theory may run
with change of energy scale. In particular, the RG for the coefficient
$a_4$ in the vacuum action (\ref{vacuum}), (\ref{HD}) has a general
form
\beq
\mu \frac{da_4}{d\mu}\,=\,\be_4\,=\,l_1 + l_2\xi + l_3 \xi^2\,,
\label{RGa4}
\eeq
where the coefficients $l_{1,2,3}$ are given by power series in
coupling constants, corresponding to the loop expansion. We assume
that the high energy GUT-like model
(supersymmetric or not) includes gauge $g$, Yukawa $h$ and
four-scalar $f$ couplings, hence  \ $l_{1,2,3}=l_{1,2,3}(g,h,f)$.

Indeed, all quantities in Eq. (\ref{RGa4}) are also running parameters
and satisfy their own RG equation. In particular, the equation for
$\xi$ has a general form
\beq
\mu \frac{d\xi}{d\mu}\,=\,\be_\xi\,=\,l_4 + l_5\xi\,,
\label{RGa4-2}
\eeq
where \ $l_{4,5}=l_{4,5}(g,h,f)$. In principle, the running
(\ref{RGa4-2}) may significantly change both sign and magnitude
of $\xi$, even at the short interval on the energy scale. The
necessary condition for this intensive running is large values
of at least some of the couplings $g,h,f$. This situation is
possible near the transition, since it can be related to
formation of condensate and then the non-perturbative
regime may take place.

Note that at one loop the expressions are much simpler,
\beq
\be_4^{(1)} &=& l_3 \Big(\xi-\frac16\Big)^2\,,
\label{RGa4-3}
\\
\be_\xi^{(1)}  &=& \Big(\xi-\frac16\Big) l_5\,,
\label{RGa4-1}
\eeq
where
\beq
l_5 & \sim & l_{51} f + l_{52} h^2 + l_{53} g^2\,.
\label{RGa4-4}
\eeq
In the last formulas the coefficients
$\,l_{3}$, $l_{51}$, $l_{52}$, $l_{53}\,$
are model-dependent constants.

At the one-loop order the conformal values $\xi=1/6$ and $a_4=0$
are fixed points, which can be stable in either
UV or IR \cite{BShYa} (many examples and further references can
be found in this work and in the book \cite{book}). However, at
higher loops the conformal value $\xi=1/6$ is not a fixed point,
as it was found for a scalar field in \cite{hath82} and can be
also established from a general considerations \cite{Bexi}.

It is natural, albeit not necessary, to assume that the value of
$\xi$ in the far UV is conformal\footnote{For instance, this is
requested by the field-particle correspondence (traceless
$T^\mu_\nu$) for the effectively massless free fields, since
we assume asymptotic freedom in the fundamental theory.}.
Suppose the ``far UV'' corresponds to the
sub-Planckian energies. Then the value of $\xi$ can become very
much different from conformal already at the GUT scale, due to
the running (\ref{RGa4-2}) in the framework of GUT theory. Hence
when it comes to the transition from GUT to some lower energy
theory, $\xi$ may be essentially away from the conformal point,
even if it was at this point in the UV. Another important
point is that around the scale of stable-unstable
transition some of the interactions may become strong.
Then, according to (\ref{RGa4-2}), the
$\be$-function for $\xi$ may be given by an infinite power
series of large couplings. Assuming that this series is
convergent, one can see that there is nothing wrong in
a very intensive running of $\xi$ on a very short interval
of the energy scale, before the masses of the fields grow
large and running of $\xi$ stops due to the IR decoupling.

The next observation is that if  $\,\left|\xi\right|\,$ becomes
very big, then the
coefficient $a_4$ can become even much greater, due to the quadratic
dependence in (\ref{RGa4}), and especially assuming large values
of couplings before the ``confinement'' of the GUT degrees of
freedom and the non-perturbative nature of Eq. (\ref{RGa4}) in
this situation. It is worth noticing that both $\xi$ and $a_4$
are not couplings in the semiclassical theory, hence there is
no contradiction to have their values large, even within the
perturbative approach. Indeed, these arguments can easily
explain the value of $\xi \approx 40,000$, which is
requested for the Higgs inflation \cite{Shaposh}. Equally
well, or even more natural, these arguments can explain that
the value of $a_4$ is about $5 \times 10^8$, which is roughly
the square of the mentioned value of $\xi$.

It may be a very interesting problem to construct a model of
GUT and its breaking into Standard Model plus a hidden Dark
sector, which yields the picture described above. But since
this consideration is beyond the scope of the present work,
let us describe the second possibility to gain a very large
value of the coefficient $a_4$ of the $R^2$-term.

\subsection{Spontaneous symmetry breaking (SSB) with non-zero $\xi$}

As a second example, we review how the $R^2$-term can emerge due
to the spontaneous symmetry breaking (SSB) in the presence of
non-minimal coupling between scalar field and scalar curvature,
$\xi R \ph^2$. Let us stress that a non-zero $\xi$ is a necessary
condition of renormalizability of the theory with the Higgs field
or its analogs in GUT models. Another question of whether the large
value of $\xi$ is ``natural''.

As we discussed above, quantum corrections can produce an
intensive  running of $\xi$. Then, since $\xi$ is a dimensionless
quantity, it can not be regarded large or small by itself. In order
to evaluate whether $\xi$ is large or not, one has to compare the
corresponding dimensional combinations with some reference quantity.
In our case the comparison should be done between $\xi R$
and the the square of mass $m$ of the scalar field, since they
always emerge in a linear combination $\mp m^2 + \xi R$ (the
choice of a sign depends on whether the SSB is assumed or
not). For instance, in the present-day universe and Higgs field
the numbers are $\,m^2 = m_H^2 \propto 10^4 GeV^2\,$ and
$\,R \sim H_0^2 \propto 10^{-84} GeV^2$ ($H_0$ is the value of
Hubble parameter). Obviously, the values of $\,\xi = 10^4 - 10^5$
do not look large in this case. Indeed, the situation may be
different in the early Universe, since the curvature has been
much greater then. In this situation the curvature effects can be
relevant for the SSB and Higgs, that is well-known from the studies
of curvature-induced phase transitions (see, e.g., \cite{book}) and
Higgs inflation \cite{Shaposh,BKS-2008,BezSha-2009}

Let us consider how the
$R^2$-term emerges in the induced action of gravity in the theory
with non-minimal interaction $\,\xi\,$ and SSB. The considerations
presented below are not directly related to the Higgs field in the
Standard Model and can be also applied to more general theories
at different energy scales.

We start by briefly reviewing SSB in curved space-time, in a way
it was originally discussed in \cite{sponta}. Consider the classical
action of a scalar field $\,\ph$,
\beq
\nonumber
S
&=&
\int d^4x\sqrt{-g}\,\Big\{
g^{\mu\nu}\,\pa_\mu \ph^*\,\pa_\nu\ph
\,+\,\mu_0^2\,\ph^*\ph\\
\,&+&\,\xi\,R\,\ph^*\ph \,-\, \la(\ph^*\ph)^2 \Big\}\,.
\label{scalar}
\eeq
The vacuum expectation value $v$ for the scalar field is defined
by the relation
\beq
-\,{\Box}v\,+\,\mu_0^2\,v\,+\,\xi R\,v\,-\,2\la v^3\,=\,0\,.
\label{SSB 2}
\eeq
For a minimal interaction case $\,\xi=0$ we have a constant
solution
\beq
v_0^2\,=\,\frac{\mu_0^2}{2\,\la}\,.
\label{SSB 3}
\eeq
However, in a generic curved space one can not find a constant
solution due to the potentially variable curvature scalar. Then
the $\Box$-term in (\ref{SSB 2}) can not be neglected and a
closed-form compact solution is impossible. At the same time,
one can obtain a solution in the form of the power series in $\xi$,
\beq
v(x)\,=\,v_0\,+\,v_1(x)\,+\,v_2(x)\,+\, \cdots \,.
\label{SSB 3.gen}
\eeq
regarding (\ref{SSB 3}) as a zero-order approximation.

In the first order we find \cite{sponta}
\beq
v_1\,=\,\frac{\xi\,v_0}{{\Box}-\mu^2+6\la v_0^2}\, R
\,=\,\frac{\xi\, v_0}{{\Box}\,+\,4\la v_0^2}\,R\,.
\label{SSB 3.1}
\eeq
It is not difficult to derive the next orders of this expansion,
for instance
\beq
\nonumber
v_2&=&\frac{\xi^2\,v_0}{\Box + 4\la v_0^2}\,
\Big\{
R\,\frac{1}{\Box + 4\la v_0^2}\,R
\\
&-&6\la v_0^2
\Big(\,\frac{1}{\Box + 4\,\la v_0^2}\,R\,\Big)^2\Big\}\,.
\label{SSB 3.2}
\eeq
In the point of the minima we meet the induced gravity action.
Replacing the solution (\ref{SSB 3.gen}) into the scalar action
(\ref{scalar}), we obtain the induced low-energy action of vacuum,
depending only on the metric,
\beq
\nonumber
S_{ind}
&=&\int d^4x\sqrt{-g}\,\Big\{\,g^{\mu\nu}
\,\pa_\mu v \,\pa_\nu v
\\
\nonumber
&+&(\mu_0^2+\xi R)\,v^2
\,-\,\la\,v^4\,\Big\}\,.
\label{SSB 3.4}
\eeq

Making an expansion in the powers of $\xi$, at the second
order we arrive at the expression
\beq
S_{ind}
&=&
\int d^4x\sqrt{-g}\,\Big\{-\,v_1{\Box}v_1
+ \xi R\,(v_0^2+2v_0v_1)
\nonumber
\\
&+&
\mu^2\,(v_0^2+2v_0v_1+2v_0v_2+v_1^2)
\nonumber
\\
&-&
\la\,(v_0^4+4v_0^3v_1+4v_0^3v_2+6v_0^2v_1^2)\Big\} +\,{\cal O}(R^3)
\nonumber
\\
&=&
\int d^4x\sqrt{-g}\,\Big\{
\,\la v_0^4\,+\,\xi Rv_0^2
\nonumber
\\
&+&\xi^2\,v_0^2\,R\,\frac{1}{{\Box}
+4\,\la v_0^2}\,R\,+\, \cdots \Big\}\,.
\label{SSB 3.5}
\eeq

The first two terms in the last action represent an
induced cosmological constant (CC) and Einstein-Hilbert
terms, correspondingly.

Let us make a few observations concerning the last result
(\ref{SSB 3.5}). The induced CC density in the first term
is huge compared to the observed value, hence the compensating
vacuum CC should be introduced. There is a extensive discussion
of the fine-tuning requested for this compensation see, e.g.,
the standard review \cite{weinberg89} and recent works treating
this problem in the QFT framework \cite{nova}.
The second term in (\ref{SSB 3.5}) is the induced Einstein-Hilbert
term. For $\,\xi v_0^2 \ll M_P^2\,$ the value of the coefficient
of this term is not sufficient to have a purely induced gravity,
hence the corresponding vacuum term is requested. In the case of
the Standard Model Higgs and $\xi \propto 10^4$ the induced term
is just a very small correction to the vacuum term. One can say
that the situation is opposite to the one with the CC term.

The third term in (\ref{SSB 3.5}) is quadratic in scalar curvature
and in $\xi$ and it is non-local. It is easy to see that this term
behaves in a very different way in the UV and IR limits. The
definition of UV here is related to the magnitude of derivatives
of the curvature tensor, compared to $v_0^2$. In the case when
$\Box R \gg v_0^2R$, the term is essentially non-local and
shows the global scaling which is identical to the one of the
Einstein-Hilbert term. Let us mention, by passing, that the next
local term, of the $R \big(\Box + 4\la v_0^2\big)^{-2}R$-type,
cancels identically, hence there is no similar correction to
the CC-term from SSB\footnote{This does not mean that these
corrections are impossible within other approaches. There was
recently an interesting work \cite{MM} (see further references
therein) about the cosmological relevance of the massless version
of such a term, proposed originally in \cite{sponta}.}.
For inflation, when $R$ is approximately constant, we can
assume an opposite relation $\Box R \ll v_0^2R$. Then the
third term in (\ref{SSB 3.5}) becomes effectively local and
equal to
\beq
S_{ind} = \frac{\xi^2}{4\la}\,\int d^4x\sqrt{-g}\,R^2\,.
\label{SSB 3.6}
\eeq
One can note that in the same approximation the next order
terms are suppressed by higher powers of $\,\xi R/v_0^2$ and
therefore the last term represents the leading quantum
contribution with higher powers of scalar curvature
being small corrections to it.

One can easily see that the term (\ref{SSB 3.6}) is
close to what we need for a successful ``jump'' in the value of
the coefficient of the $R^2$-term due to the phase transition
related to SSB. Assuming that $\xi$ has a large value, e.g., due
to the mechanism which we discussed in the previous subsection,
the small value of the four-scalar coupling $\la$ in the IR is
enhancing the effect of running.

Different from the running
of $a_4$, the induced value (\ref{SSB 3.6}) has a definite
positive sign. According to Eqs. (\ref{c}), (\ref{condition})
and (\ref{quantum1}) this is the same sign which we need for
the unstable $R+R^2$ model \cite{star,star-a}.
In the GUT-like models with several scalars
there are typically different $\la$ and $\xi$ for each of these
scalars. Then it is sufficient that one of the combinations
$\,\xi^2/\la\,$ becomes very large at the instant when the
scalar is freezing in the vacuum state, to provide a desired
huge value of induced coefficient $a_4$. It would be definitely
interesting to construct an explicit realization of this
situation in the framework of some GUT-like model.


Unfortunately the consideration presented above does not work
for the $R^2$-inflation \cite{star}, because during inflation the
magnitude of the product $\xi R$ is too large and the
expansion (\ref{SSB 3.gen}) is not appropriate\footnote{We
are grateful to the anonymous referee for indicating this
important point to us.}. One of the possibilities is to consider a
different expansion using de Sitter starting point
${\mu_0^\prime}^2 = \mu_0^2 + \xi R$ instead of $\mu_0^2$.
However, this approach also meets a serious problem.

It is worthwhile to discuss the situation in details and see if the
appearance of the large $R^2$ term can be explained by  the
modified SSB for a sufficiently low value of the Ricci scalar  $R$.
This value should be still sufficient for the unstable $R^2$
inflation to occur with the parameter $a_4$ following from
observational data. For a slow-roll quasi-de Sitter inflation,
when $R<0$ can be considered as a constant in the zero
approximation, the non-trivial solution of  Eq.
(\ref{SSB 2}) is
\beq
{v_0^\prime}^2=\frac{\mu_0^2+\xi R}{\la}\,.
\label{SSBphi}
\eeq
Thus, if we want to have symmetry restored at large $|R|$
and $R<0$, we have to take $\xi>0$,
i.e., the same sign of $\xi$ as in the case of conformal coupling
\ $\xi_{{\rm conf}}=1/6$ \ and opposite to that for the Higgs
inflationary model \cite{Shaposh}.
Then the SSB occurs for $|R|<\mu_0^2/\xi$, and the value of
${v_0^\prime}$ is given by Eq. (\ref{SSBphi}).

Replacing Eq. (\ref{SSBphi}) into the action (\ref{scalar})
and adding the Einstein term, we get the effective
Lagrangian density in the quasi-static (slow-roll) case:
\beq
L=-\frac{{M^\prime_P}^2 R}{2}+\frac{\xi\mu_0^2 R}{2\lambda}+\frac{\xi^2 R^2}{4\lambda}
\label{efflagr}
\eeq
where $M^\prime_P = (8\pi G)^{-1/2}$ is the reduced Planck mass.
In this model, the unstable
$R^2$ inflation occurs for values of scalar curvature
$|R|\gg R_I=\la {M^\prime_P}^2/\xi^2$.
Thus, $\la/\xi^2$ should be small to justify quasi-classical
description of space-time. This condition can be easily achieved. However, the condition that the SSB occurs at curvatures
$|R|\gg R_I$ requires also
\beq
\mu_0^2\gg \xi R_I
&=&
\frac{\la {M^\prime_P}^2}{\xi}\,.
\label{ef}
\eeq
Then the coefficient in front of $R$ in Eq. (\ref{efflagr}) has
the wrong, positive sign (i.e. gravity become repulsive at low
curvature, in particular, in almost flat space-time). Therefore,
we come to conclusion that  it is not possible to use this type
of SSB to generate the large coefficient $a_4$ needed for
viable unstable inflation. At the same time, there is a chance
that some other modified scheme related to phase transition
may be working.

\subsection{Unstable phase with a large $R^2$-term}

Assuming that there is a desired increase of the value of $a_4$ in the
transition period, it is natural to ask how it will change the evolution
if the universe in the consequent unstable period of inflation. In order
to see this, we repeated the analysis of Sects. 3 and 4, but this time
with a very large value of $a_4 = 5\times 10^8$.

Qualitatively, the
phase diagram does not change too much, since the system is still in
the ``right'' part of the phase plane of Fig. \ref{fig7}. The plot
for the trajectory is similar to the one in Fig. \ref{fig8}, but
now much closer to the origin of coordinates $x$ and $y$.

\begin{figure}[!h]
\centering
\includegraphics[height= 6.0 cm,width=\columnwidth]{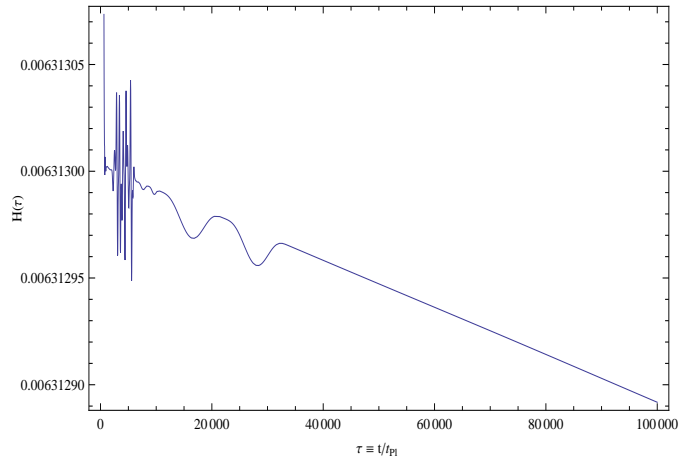}
\caption{The plot of $H(t)$ for the theory with one photon
and $a_4 = 5\times 10^8$.}
\label{figN}
\end{figure}

The result of numerical analysis is shown in Fig. \ref{figN}.
It is easy to see that the dynamical system of our interest suffers
from initial oscillations which last a very short period.
After that the dependence looks linear.
\\

The interpretation of the plot in Fig. \ref{figN} is straightforward.
The initial very fast change of $H(\tau)$ represents a remnant of the
stable phase, with a relatively large initial value of $H$. After that
the huge $R^2$-term starts to dominate. After some oscillations the
Universe starts the period of unstable inflation in the $R+R^2$-model.
The analytic expression for this phase is very similar to Eq.
(\ref{parabola}),
\beq
\si(t) &=& H_1\,t \,-\, \frac{M^2}{12}\,t^2
\,+\, {\cal O} \big(\ln (t_f-t)\big)\,,
\label{laph83}
\eeq
where $t_f$ corresponds to the end of inflation,
$H_1$ is an integration constant and $M \ll H$ in the given
phase \cite{star83}. Both formulas (\ref{parabola}) and (\ref{laph83})
lead to the approximately linear time dependence of $H$ which we
can also observe in the plot under discussion.

It looks like the transition from stable to unstable phases is rather
successful in the presence of a huge $R^2$-term, since it leads to the
known dynamics after this transition, and this is exactly the dynamics
which passed some tests in comparison with observational data. After
all, the distinguished feature of the inflationary model based on
the stable/unstable transition is the presence of preliminary stable
phase. The consequences may be not observable, but this phase provides
right initial conditions for the physically testable unstable phase.

\section{Conclusions}
\label{Sect6}

The anomaly-driven inflation \cite{star,star-a} can be stable or unstable
depending on the particle content of the underlying quantum field
theory. In the course of inflationary expansion the number of
the ``active'' fields can change, especially due to the quantum
decoupling of heavy particles from gravity \cite{apco}.
For example, one can expect the transition
from stable to unstable phases in the supersymmetric versions of
the Standard Model or GUTs, due to the decoupling of the $s$-particles.

The detailed description of the transition period is far beyond the
available theoretical methods of quantum field theory, because this
situation requires the description of the vacuum quantum effects
in the case when Hubble parameter is of the same order of
magnitude as the mass of the quantum massive field. For this reason,
we use the most simple phenomenological approach to this problem, by
assuming that the unstable part starts exactly in the point when the
stable phase ends. The unstable particle contents may lead to the
graceful exit to the usual radiation-dominated evolution, or to a
very violent ``hyperinflation''-like behavior \cite{star,vile,anju}.
By using numerical methods and also by stitching the solutions on
the phase diagrams of the theory we have found that the point when
the stable evolution ends exactly corresponds to the initial data
of the desirable type of unstable evolution, such that the
``hyperinflation''-type solutions are ruled out.

In order to have a phenomenologically successful $R+R^2$ inflation
one need to explain also a relatively large value of the
coefficient $a_4$ of the $R^2$-term. We discussed this issue
starting from the renormalization group running of the non-minimal
parameter $\xi$ and using two alternative ways to generate a huge
$R^2$-term. The conclusion is that both mechanisms, namely the
renormalization group running of vacuum $a_4$ and the SSB-based
induced gravity, are capable to provide the coefficient of the
$R^2$-term
in the desirable range. The most important ingredient in both
cases is a large value of the non-minimal parameter
$\xi \propto 10^4$. From the QFT side, this means that it would
be interesting to design the field theory models which could
provide an intensive running of $\xi$ from the UV to IR, in
either perturbative or non-perturbative frameworks. The first
step in the non-perturbative direction has been done recently
in \cite{XiFRG}, and the considerations presented above show
that this is a phenomenologically relevant subject.

Qualitatively, the output of our work means that the transition
from stable to unstable version of anomaly-driven inflation can
occur successfully, at least it is so within the sharp cut-off
approximation. In this case no special conditions for the initial
data are requested in the model. All these statements correspond to
the dynamics of the conformal factor of the metric. It would be
very interesting to extend it further to the case if initially
anisotropic metrics. There is a strong expectation that
anisotropy disappears rapidly during the stable phase, but this
feature still requires a detailed investigation.

\section*{Acknowledgments}
T.P. is grateful to CAPES for supporting his Ph.D. project.
I.Sh. was partially supported by CNPq, FAPEMIG and ICTP.
The work of A.S. in Russia was supported by the RSF grant
16-12-10401.
His visit to the Utrecht University was supported by the Delta
ITP grant BN.000396.1. He thanks Profs. S.~Vandoren and
T.~Prokopec and Drs. J. van Zee for hospitality during the visit.



\end{document}